\documentclass[twocolumn,prb,showpacs]{revtex4}
\usepackage{graphicx}
\usepackage{color}
\usepackage{dcolumn}

\usepackage{tabularx}

\usepackage{float}

\usepackage{amsmath}
\newcommand{\comment}[1]{}

\begin{document}


\title{Optical conductivity spectra of the rattling phonons and charge carriers in type-VIII clathrate Ba$_8$Ga$_{16}$Sn$_{30}$}
\author{K. Iwamoto$^1$, T. Mori$^{1,2}$, S. Kajitani$^1$, H. Matsumoto$^1$, and N. Toyota$^1$}
\affiliation{$^1$Department of Physics, Graduate School of Science, Tohoku University, Sendai 980-8578, Japan}
\affiliation{$^2$Division of Materials Science, University of Tsukuba, Tsukuba 305-8573, Japan}
\author{K. Suekuni$^3$, M. A. Avila$^{3,5}$, Y. Saiga$^3$, and T. Takabatake$^{3,4}$}
\affiliation{$^3$Department of Quantum Matter, ADSM, $^4$Institute for Advanced Material Research, Hiroshima University, Higashi-Hiroshima 739-8530, Japan}
\affiliation{$^5$CCNH, Universidade Federal do ABC (UFABC), Santo Andr\'e, SP, 09210-580, Brazil}

\date{\today}
\begin{abstract}
We have investigated optical conductivity spectra of $n$- and $p$-type Ba$_8$Ga $_{16}$Sn$_{30}$ ($\alpha$-BGS) with type-VIII clathrate structure, at temperatures from 296\,K down to 6\,K with a terahertz time-domain spectrometer (0.2\,-\,2.5\,THz). 
The continuous spectra contributed from charge carriers are dispersive in this frequency range and also temperature- and carrier type-dependent. 
The Drude-Smith model taking multiple-scatterings of charge carriers into account well reproduces those data. 
The relaxation rate of the $n$-type carriers decreases more sharply than that in the $p$-type material, suggesting that a stronger electron-phonon interaction may exist in the $n$-type than in the $p$-type. 
On the other hand, the localized infrared-active modes observed at 1.3\,THz and 1.7\,THz, identified as the rattling phonons of the Ba$^{2+}$ ion's quasi-on-center vibrations, become soft and broad significantly with decreasing temperature as well as observed in type-I BGS and BGG (Ba$_8$Ga$_{16}$Ge$_{30}$) clathrates. 
The softening in the $n$-type is smaller by about 30\,\% than in the $p$-type, whereas the linewidth brodening is almost the same independently on the carrier type. 
The difference in the softening is discussed with a scenario where the interaction of rattling phonons with carriers can modify the anharmonic potential of the guest ions. 
The anomalous broadening at low temepratures is also discussed by the impurity-scattering model presented for a rattling-phonon system strongly hybridized with acoustic cage phonons. 
\end{abstract}


\maketitle

\section{Introduction}

Cagelike materials, such as clathrate compounds, \cite{Nolas1998,Cohn1999,Sales2001} often exhibit metallic electrical conductivity but a heavily suppressed thermal conductivity resembling glassy materials, which can lead to their potential application in thermoelectric devices. 
In general, they have a network structure of atomic cages encapsulating a guest ion.
These guest ions are loosely bounded by the local potential of an electronegative cage and therefore vibrate with a large amplitude.
Due to its low frequency, such a localized vibrational mode of the guest in the oversized cage (referred to as {\it rattling phonon}) hybridizes and interferes strongly with the heat-carrying acoustic phonons of the cage network.\cite{Christensen2008}
Thus, the rattling phonon is expected to play a key role in the above-mentioned characteristic transport phenomena.

\begin{figure}[!tb]
\includegraphics[width=0.28\textwidth ]{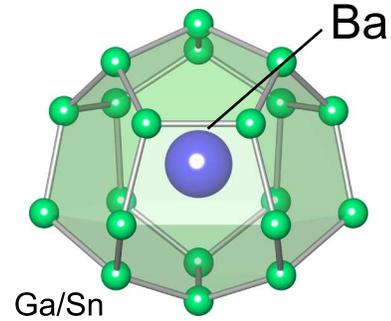}
\caption{(color online)
Distorted dodecahedral cage of $\alpha$-BGS. 
}
\label{fig1}
\end{figure}

The type-VIII clathrate Ba$_8$Ga$_{16}$Sn$_{30}$, hereafter abbreviated as $\alpha$-BGS, adopts a cubic structure with space group $I\bar{4}3m$, presenting 8 distorted dodecahedral cages formed by a network of Ga and Sn.\cite{Eisen1986,Huo2005,Avila2006c,Avila2006b}
It is a so-called Zintl compound, stabilized by charge compensation, that each cage is donated 2 electrons from a guest Ba$^{2+}$.
Figure \ref{fig1} illustrates a disordered dodecahedral cage and the guest Ba$^{2+}$ ion.
A fine tuning of the Ga/Sn relative concentration slightly off the stoichiometry can turn the system into a heavily doped semiconductor, and hence control both the charge carrier sign and density. \cite{Avila2006c,Avila2006a, Avila2008, Suekuni2008}

One of the most striking features of $\alpha$-BGS is the fact that the lattice thermal conductivity $\kappa _L$ behaves quite differently depending on the sign of charge carriers.\cite{Avila2006b,Suekuni2008}
In the $p$-type materials $\kappa_L$ shows an almost glasslike temperature dependence akin to disordered materials, with a plateau around 10-30\,K, whereas in $n$-type ones $\kappa_L$ has a normal crystalline peak around 10\,K.
Such behaviors in $\kappa _L$ have also been observed in a few type-I clathrates, Ba$_8$Ga$_{16}$Ge$_{30}$ (BGG)\cite{Avila2006a,Avila2006b,Bentien2004} and Ba$_8$Ni$_{6-x}$Ge$_{40+x}$ (BNG) \cite{Bentien2006}, where the carrier types have been found fully tunable.
It is noteworthy that so far only Ba-based clathrates (but not all of them) have shown such full tunability and strong carrier-dependent transport. 
Although it is expected that the coupling of the heat-conducting acoustic phonons with the guest ion rattling is {stronger} in $p$-type materials \cite{Avila2006b}, the origin of the carrier-dependent $\kappa_L$ in $\alpha$-BGS and others remains an intriguing open question, despite the fact that the detailed interaction between rattling and acoustic phonons has been intensively studied and debated in literature.

In a recent paper, we have reported the optical conductivity spectra of the infrared-active rattling phonons by using terahertz time-domain spectroscopy (THz-TDS) in the type-I clathrate BGG,\cite{Mori2009,Iwamoto2013} whose carrier-dependence in $\kappa_L$ is quite similar to that of $\alpha$-BGS. \cite{Avila2006b}
For $p$-type BGG, where any effects of charge carriers are negligible due to the quite low carrier density of the measured samples, the lowest-lying vibrational mode of the Ba$^{2+}$ guest ion consists of a sharp Lorentzian peak at 1.2\,THz superimposed on a broad tail weighted in the lower frequency range around 1.0\,THz. \cite{Iwamoto2013}
With decreasing temperature, an unexpected linewidth broadening of the phonon peak is observed, together with monotonic softening of the phonon peak and an enhancement of the tail structure.
From the analysis based on an impurity scattering model for rattling modes that are strongly hybridized with cage acoustic phonons, it is suggested that the anomalous features of the rattling phonon spectra arise from both the softening due to anharmonicity and the strong hybridization between rattling and acoustic phonons.
In BGG, certain dissipative mechanisms into a continuum of multi-bosonic particle states, such as phonon-phonon scattering, may not have any effects on the rattling phonons at the $\Gamma$ point of the first Brillouin zone.
The rattling excitation in the single-well, or {\it on-center}, potential may have a more propagating nature through the hybridization with the acoustic phonon, and the local potential may contribute rather as a mean field. \cite{Dahm2007, Iwamoto2013}

The optical conductivity spectra of the infrared-active rattling phonons have also been investigated in type-I Ba$_8$Ga$_{16}$Sn$_{30}$ ($\beta$-BGS) \cite{Mori2011} in which $\kappa_L$ is strongly glasslike irrespective of the charge carriers.
With decreasing temperature, a single broad peak of the lowest-lying Ba$^{2+}$ guest mode becomes split into two subpeaks and also shows a linewidth broadening.
The splitting is understood by assuming a multi-well {\it off-center} anharmonic potential, \cite{Matsumoto2009} revealed by structural analyses \cite{Suekuni2008}.
However, the cause of broadening remains an open question due to the following:
In $\beta$-BGS, according to the impurity-scattering model for the hybridized phonon system, \cite{Iwamoto2013} a crossover from a mean field regime to a degenerate level transition \cite{Matsumoto2009} is expected with decreasing temperature, since a strong hybridization of {\it off-center} rattling with acoustic phonons is predicted theoretically. \cite{Nakayama2011}
Since those samples have significant carrier densities of order $10^{20}$\,cm$^{-3}$ even at low temperature, \cite{Suekuni2008} the interaction with charge carriers cannot be neglected.

In such a context, the relationships between the charge carriers and the phonon system also remains as one of the major issues required to clarify the transport mechanisms in cage systems.
In type-I clathrates, phonon scattering contributions from localized carriers have been suggested as a relevant origin for carrier-dependent $\kappa_L$, \cite{Bentien2004,Bentien2006} as such effects may change the dynamical properties of the phonon system depending on the density and sign of the charge carriers.
Measurements that probe the dynamical properties of both charge carriers and phonons should thus be important.
Although the optical conductivity spectra of the type-VIII clathrate Eu$_8$Ga$_{16}$Ge$_{30}$ ($\alpha$-EGG) have also been reported, \cite{Sichelschmidt2005} there are few discussions about the lifetimes of charge carriers and phonons resulting from electron-phonon interaction.
As supported by structural analysis, \cite{Avila2006b} neutron magnetic resonance, \cite{Tou2011} and ultrasonic measurements, \cite{Ishii2005} the rattling phonons of Ba$^{2+}$ in $\alpha$-BGS can be approximated as on-center, \cite{Avila2006b,Suekuni2008} and are therefore much simpler than those in $\beta$-BGS, to analyze the relationships between the charge carriers and the phonon system.
The dc resistivity also shows metallic behavior for both types of charge carriers, \cite{Suekuni2008} indicating that there is sufficient carrier density for their interaction with rattling phonons to be expected even at low temperatures.
Since the magnitude of the dc resistivity is about 5-20\,${\rm m} \Omega {\rm cm}$, \cite{Suekuni2008} it is expected that the contribution from rattling phonons in the optical conductivity is comparable to that from charge carriers.
For the above reasons, $\alpha$-BGS is one of the best materials to study the dynamical properties of both charge carriers and rattling phonons and hence the electron-phonon interaction in clathrate compounds.

In this paper, we report a thorough investigation on the optical conductivity spectra of $\alpha$-BGS doped with both $n$- and $p$-type carriers.
The data show clear carrier dependence between $n$-$\alpha$-BGS and $p$-$\alpha$-BGS, which appears particularly in the relaxation rates of conducting carriers and in the frequencies of the Ba$^{2+}$ guest ion vibrations.
We discuss these dependencies in terms of impurity scattering in the hybridized phonon system and of electron-phonon interactions.
The carrier-type dependence observed in the optical conductivity can be understood by taking into account the interaction between charge carriers on Ba$^{2+}$ ions and the rattling vibration of Ba$^{2+}$ ions.

\section{Experiment}

\begin{figure}[!tb]
\includegraphics[width=0.48\textwidth ]{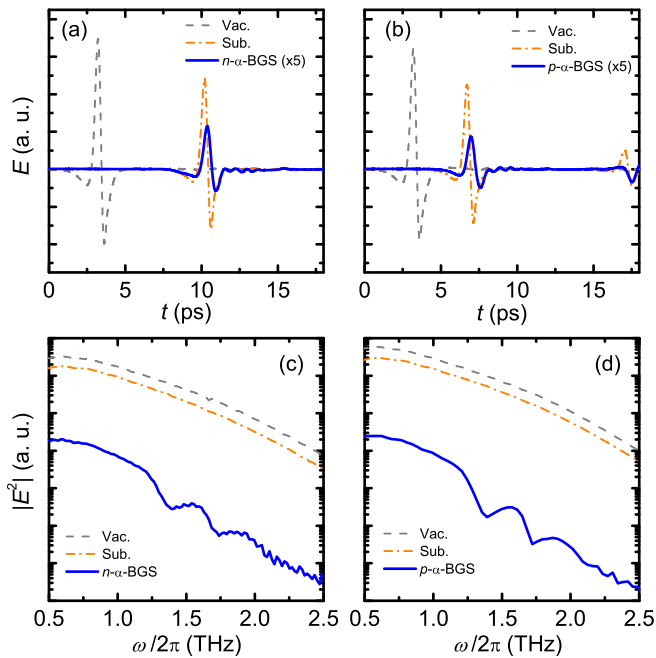}
\caption{(color online)
(a, b) Time evolution of THz-wave electric fields for the type-VIII Ba$_8$Ga$_{16}$Sn$_{30}$ ($\alpha$-BGS) doped with (a) $n$- and (b) $p$-type carriers.
These are obtained from three different sets of measurements:
vacuum (Vac., dashed lines), substrate only (Sub., dash-dotted lines), and $\alpha$-BGS sample glued onto the substrate (solid lines).
(c, d) Intensity spectra of THz-wave electric fields where the time evolution are converted for (c) $n$-type and (d) $p$-type $\alpha$-BGS. }
\label{fig2}
\end{figure}

Single crystals of $\alpha$-BGS doped with $n$- and $p$-type carriers were grown by a self-flux method. \cite{Huo2005,Avila2006c}
To obtain sufficient transmitting signals through the conductive sample, single crystal disks of 2\,-\,5\,mm in diameter were attached onto a sapphire substrate and polished down to about 10\,-\,15\,$\mu$m in thickness by using diamond abrasive films.
We used sapphire substrates with thickness of 1.0\,mm for $n$-$\alpha$-BGS and 0.5\,mm $p$-$\alpha$-BGS.
The frequency resolutions are estimated as about 0.05\,THz and 0.1\,THz for $n$- and $p$-$\alpha$-BGS, respectively.
We note that these values, determined by the thickness of the substrate, \cite{Iwamoto2013} are sufficiently narrow to allow analysis of the dynamical properties of guest modes.

The present measurements covering a frequency range of 0.5 - 2.5\,THz were carried out with a commercial spectrometer (RT-20000, Tochigi Nikon Co. Ltd) which uses a standard technique for the transmission configuration, and the temperature ranged from 6\,K to room temperature. \cite{Mori2008, Iwamoto2013, Mori2009, Mori2011}
The measured time evolution of the transmitting electric fields through vacuum (Vac.), substrate only (Sub.), and sample glued onto the substrate, shown in Fig. \ref{fig2} (a) and (b), are converted by Fourier transform into the intensity spectra and the phase of electric fields, shown in Fig \ref{fig2} (c) and (d).
Then the complex refractive indexes $\hat{n} (\omega) = n + i \kappa$ are numerically determined by using the following equation which takes into account multiple reflections at the surface and boundary among the sample, glue, and substrate:
\begin{equation}
\begin{split}
t (\omega) = & \frac{t_{\it vf} t_{\it fg} t_{\it gs}} {t_{\it vs}} \exp \{ i [\phi_{\it f} + \phi_{\it g} - \omega (d_f + d_g)/c_0] \} \\
& \times \{1 - r_{\it fg} r_{\it fv} \exp (2i\phi_{\it f}) - r_{\it gs} r_{\it gf} \exp (2i \phi_{\it g}) \\
& - r_{\it gs} r_{\it fv} \exp (2i \phi_{\it f} + 2i \phi_{\it g} ) \}^{-1}, \\
\end{split}
\label{eqnt}
\end{equation}
where $t_{ij} = \frac{2\hat{n}_j}{\hat{n}_i + \hat{n}_j}$ and $r_{ij} = \frac{\hat{n}_i - \hat{n}_j}{\hat{n}_i + \hat{n}_j}$ $(i,j = v,\,s,\,g,\,f)$ are complex transmission and reflection coefficients, $c_0$ is the velocity of light, $\phi_i = \hat{n}_i \omega d_i/c_0$ is the phaseshift, and the thickness $d_i$ of each medium is taken as $d_f \simeq d_g \ll d_s$.
The subscripts $v,\,s,\,g,\,f$ in Eq. \ref{eqnt} represent vacuum, substrate, glue, and sample respectively.
For numerical analyses of $\hat n_f$, we assumed a constant value for $\hat n_m = 1.66$, which was obtained from previous THz measurements.
The validity of the obtained spectra is confirmed by using the Kramers-Kronig relations between refractive index $n$ and extinction coefficient $\kappa$. \cite{Dressel}
Then, the complex conductivity $\hat \sigma$ is obtained from
\begin{equation}
\hat n_f^2 = 1 + \frac{4 \pi i \hat \sigma }{ \omega }.
\label{eqns}
\end{equation}

To note finally in this section, these spectral measurements and analyses mentioned above were thoroughly made on several samples of the optimized thickness in order to obtain the reproducible results, which will be described in next section. 

\section{Results}

\subsection{Optical conductivity spectra}

\begin{figure}[!tb]
\includegraphics[width=7.2cm]{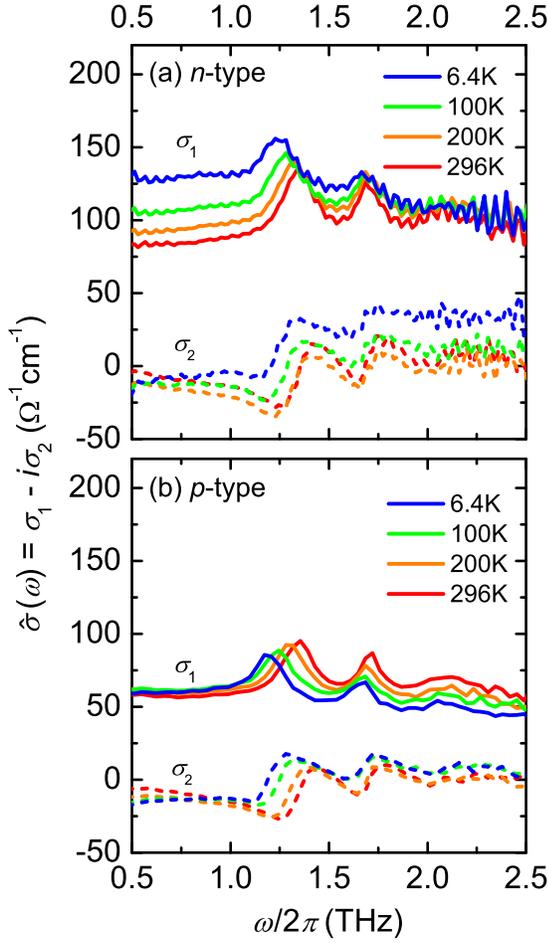}
\caption{(color online)
Temperature-dependent complex conductivity spectra $\hat{\sigma} (\omega) = \sigma_{1}(\omega) - i \sigma_2 (\omega)$ of $\alpha$-BGS doped with (a) $n$- and (b) $p$-type carriers.
Solid and dashed lines indicate the real and imaginary parts of $\hat{\sigma} (\omega)$, respectively.
Note that the horizontal axis is frequency (= $\omega/2\pi$) in\,THz.
For convenience: 1\,THz = 33.3\,cm$^{-1}$ = 48\,K = 4.14\,meV.
}
\label{fig3}
\end{figure}

Figure \ref{fig3} (a) and (b) show the real and imaginary parts of the complex conductivity spectra $\hat{\sigma} (\omega) = \sigma_{1}(\omega) - i \sigma_2 (\omega)$ from room temperature down to 6\,K for $n$- and $p$-$\alpha$-BGS, respectively.
The real part $\sigma_1(\omega)$ features two large and distinct peaks superimposed on the contributions from charge carriers, whereas the imaginary part $\sigma_2 (\omega)$ shows the corresponding derivatives with angular frequency $\omega$.
$\sigma_1$ also shows a small band around 2.2\,THz, which is almost temperature-independent.
The magnitude of the carrier contribution at room temperature is about 80 $\Omega ^{-1} {\rm cm}^{-1}$ and 60 $\Omega ^{-1} {\rm cm}^{-1}$ for $n$- and $p$-$\alpha$-BGS, respectively, which agree with the dc resistivity. \cite{Suekuni2008}
Upon cooling, the carrier contributions in $\sigma_1$ for $n$-$\alpha$-BGS monotonically increase and become dispersive, whereas for $p$-$\alpha$-BGS they only show weak temperature dependence.
The overall tilts of the total spectra in $\sigma_2$ correspond to the superpositions of an inductive response of charge carriers and a dielectric response of valence electrons.

\begin{figure}[!tb]
\includegraphics[width=6.8cm]{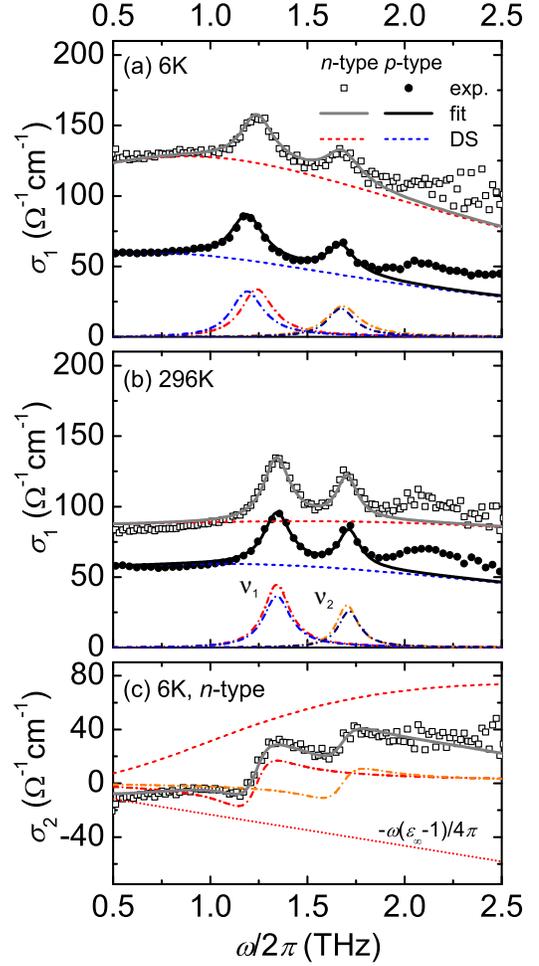}
\caption{(color online)
(a, b) Experimental data and fitting curves of the real part $\sigma_1$ of complex conductivity (a) at 6\,K and (b) at 296\,K.
Open squares and filled circles indicate the experimental data of $n$-$\alpha$-BGS and of $n$-$\alpha$-BGS, respectively,
and lines the fitting results.
Each component of the fitting curves, the contribution of charge carriers (DS, dashed lines) and those of infrared-active phonons (dash-dotted lines), is also plotted.
(c) Experimental results (open squares) and fitting curves (lines) of the imaginary part $\sigma_2$ of the complex conductivity at 6\,K for $n$-$\alpha$-BGS.
The dotted line represents the contribution of the cage electrons.
Details are described in the text.
}
\label{fig4}
\end{figure}

We then adopt a phenomenological formula for $\tilde \sigma (\omega)$ in order to separate contributions from the charge carriers and rattling phonons.
The observed carrier contribution cannot be reproduced by the conventional Drude form.
In fact, within the frequency range shown in Fig. \ref{fig3}, it can be seen that the carrier contribution grows with increasing frequencies, particularly in the low frequency region below 1\,THz, where the $\sigma _1$ response slopes up gradually although no contribution from the infrared-active phonons are expected.
Moreover, in the $n$-$\alpha$-BGS at room temperature, the gradual increase extends to higher frequency, $\sigma _1 \sim 80\, \Omega ^{-1} {\rm cm}^{-1}$ at 0.5\,THz and $\sigma _1 \sim 95\, \Omega ^{-1} {\rm cm}^{-1}$ at 2.5\,THz.
Such a behavior is also observed in other clathrate materials. \cite{Iwamoto2013,Mori2011,Mori2009,Sichelschmidt2005}.
Thus, we write the fitting expression for $\hat{\sigma}_{\it fit} (\omega)$ as
\begin{equation}
\begin{split}
\hat{\sigma}_{\it fit} (\omega) = & \frac{{\omega _p}^2/4\pi}{\gamma + i \omega} \left( 1 + \frac{\chi}{1 + i \omega / \gamma} \right) \\
 & + \sum_{j = 1, 2} \frac{(2/\pi) i \omega S_{j}}{\omega_{j}^2-\omega ^2 + i \omega \Gamma_{j}}
- \frac{(\epsilon _{\infty} - 1) \omega}{4 \pi i}.
\label{fitting}
\end{split}
\end{equation}
$\hat{\sigma}_{\it fit} (\omega)$ can be understood as a superposition of three components.
The first term indicates the Drude-Smith (DS) type contribution $\hat{\sigma} _{DS}$ of charge carriers. \cite{Smith2001}
$\omega _p$ is the plasma frequency and $\gamma$ is the relaxation rate.
The coefficient $\chi$ is $\langle \cos \theta \rangle$, the expectation value of $\cos \theta$ where $\theta$ is the scattering angle by elastic collisions. \cite{Smith2001}
For each phonon contribution, we apply the superpositions of two Lorentz oscillators $\hat L _j = L_{1j} - iL_{2j}$ $(j = 1,2)$ in the second term.
Here, $S_j$, $\omega_{j}$, and $\Gamma_j$ represent the oscillator strength, the peak frequency, and the linewidth of the $\nu_j$ mode, respectively.
$S_j$ is derived from the optical sum rule for each $\nu_j$ mode and thus defined by
\begin{equation}
S_j = \int_0 ^{\infty} d \omega L _{1j} ( \omega ) = \frac{\pi}{2} \frac{N_j q_j^2}{M_j},
\label{optsumph}
\end{equation}
where $N_j$, $q_j$, and $M_j$ are the mass, density, and charge of atoms involved in the $\nu_i$ phonon mode, respectively.
The dielectric response of the cage electrons are also taken into account as the last term, where $\epsilon _{\infty}$ is the optical dielectric constant.

Figure \ref{fig4} shows a comparison of the fitting curves (solid lines) with the experimental results (open and closed symbols) at 6\,K and 296\,K for both samples.
The individual components in Eq. \ref{fitting} are also plotted as dashed, dash-dotted, and dotted lines, which represent $\hat{\sigma} _{DS}$, $\hat L _j$, and the dielectric response of the cage electrons, respectively.
The fittings are carried out for both $\sigma _1$ and $\sigma _2$, and reproduce the experimental results very well from 0.5\,THz to 2.0\,THz.
Due to insufficient signal/noise ratio above 2.0\,THz, particularly at low temperatures in $n$-$\alpha$-BGS, we will not discuss the small band around 2.2\,THz, which may be assigned as the superpositions of several cage modes.

\subsection{Charge carriers}

\begin{figure}[!tb]
\includegraphics[width=6.8cm]{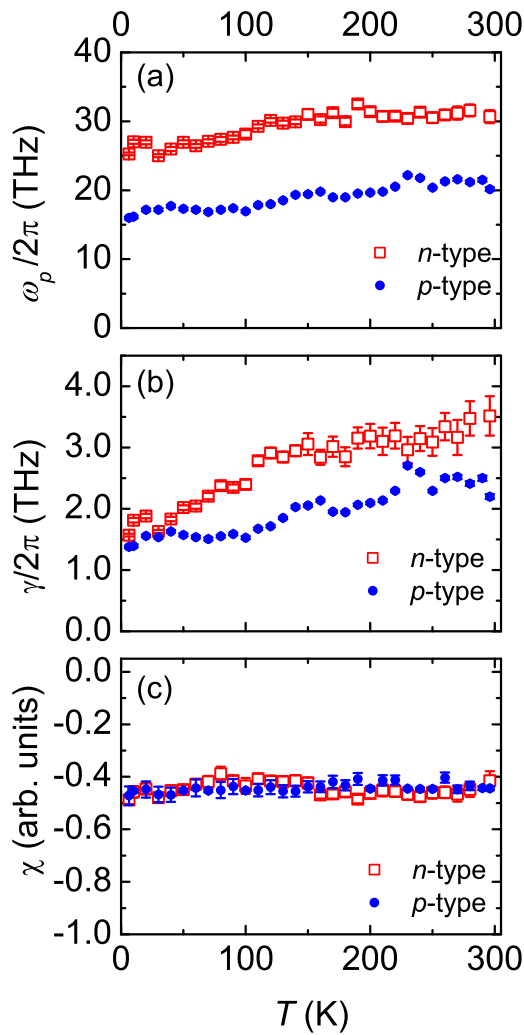}
\caption{(color online)
Temperature dependence of (a) the plasma frequency $\omega_{p}$, (b) the relaxation rate $\gamma$ of the charge carriers, and (c) the expectation value of the cosine of the scattering angle $\chi$. Open squares and filled circles represent $n$-$\alpha$-BGS and $p$-$\alpha$-BGS, respectively.
}
\label{fig5}
\end{figure}

Figure \ref{fig5} (a), (b) and (c) show the temperature dependences of the fitting parameters for the charge carriers: the plasma frequency $\omega _p$, relaxation rates $\gamma$, and $\chi$, respectively.
$\omega_p$ is connected to the charge carrier density $n_{cc}$ through the optical sum rule
\begin{equation}
\frac{\omega _p ^2}{8} = \frac{\pi}{2} \frac{n_{cc} e^2}{m^{\ast}} = \int _0 ^{\infty} d \omega \sigma _{1,cc} (\omega),
\label{optsum2}
\end{equation}
where $m^{\ast}$ is the effective mass and $e$ is the elementary charge.
As shown in Fig. \ref{fig5} (a), $\omega_p$ is about 32\,THz in $n$-$\alpha$-BGS and about 22\,THz in $p$-$\alpha$-BGS at room temperature, from which, assuming Eq. \ref{optsum2} and $m^{\ast}$ as the free electron mass $m_0$, $n_{cc}$ are estimated as about $1.5 \times 10^{19} /{\rm cm}^3$ for $n$-$\alpha$-BGS and $7.7 \times 10^{18} /{\rm cm}^3$ for $p$-$\alpha$-BGS.
Upon cooling, $\omega _p$ decreases by about 20\,\% and 40\,\% in $n$- and $p$-$\alpha$-BGS, respectively.
Such a temperature dependence may be related to charge carrier excitations from donor (acceptor) levels.
Fig. \ref{fig5} (b) shows that upon cooling the relaxation rates $\gamma$ are monotonically reduced, from about 3.5\,THz in $n$-$\alpha$-BGS and 2.5\,THz in $p$-$\alpha$-BGS at room temperature to almost half at 6\,K.
Note that $\gamma$ in $n$-$\alpha$-BGS is significantly larger than that in $p$-$\alpha$-BGS, on average by about 60\,{\%}.
Fig. \ref{fig5} (c) demonstrates that $\chi$ takes a constant value of about -0.44, irrespective of temperature or charge carrier sign.

\begin{figure}[!tb]
\includegraphics[width=6.8cm]{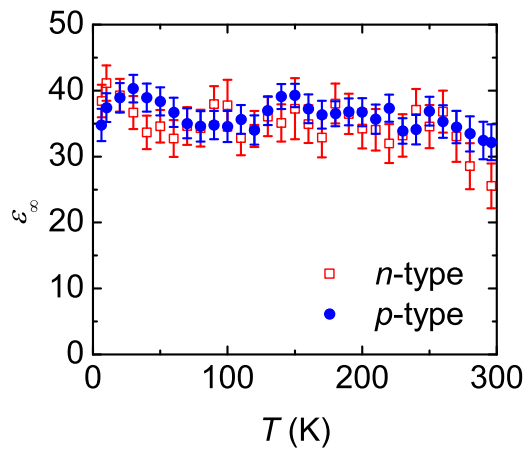}
\caption{(color online)
Temperature dependence of the optical dielectric constant $\varepsilon_{\infty}$. Open squares and filled circles represent $n$-$\alpha$-BGS and $p$-$\alpha$-BGS, respectively.
}
\label{fig6}
\end{figure}

Figure \ref{fig6} shows the optical dielectric constant $\varepsilon _{\infty}$, which remains at about 35 and is mostly independent of temperature or charge carrier sign in the measured range.

\subsection{Rattling phonons}

\begin{figure}[!tb]
\includegraphics[width=7.0cm]{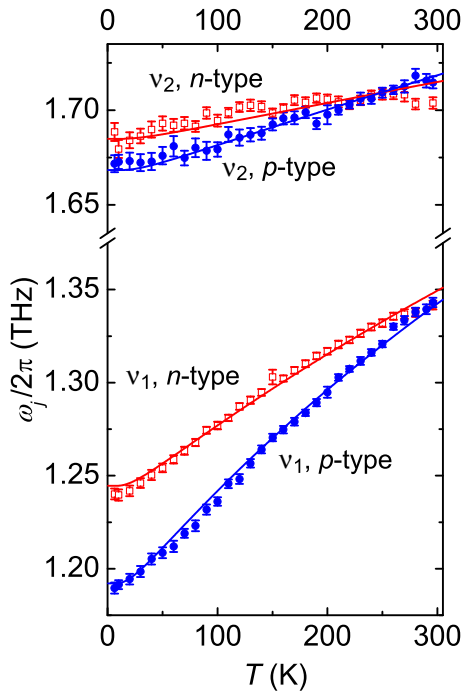}
\caption{(color online)
Temperature dependence of the peak frequency $\omega_{1}$ for the $\nu_1$ mode (lower panel) and $\omega_{2}$ for the $\nu_2$ mode (upper panel) of $\alpha$-BGS.
Open squares and open circles indicate $n$-$\alpha$-BGS and $p$-$\alpha$-BGS, respectively. The theoretical curves obtained from the anharmonic potential model \cite{Dahm2007} are also plotted as solid lines.
}
\label{fig7}
\end{figure}

\begin{table}[!b]
\caption{
Calculated phonon frequencies and fitting parameters of $\nu_1$ and $\nu_2$ modes in $\alpha$-BGS.
}
\label{tab1}
\begin{center}
{\renewcommand\arraystretch{1.2}
\setlength{\arrayrulewidth}{0.2pt}
\begin{tabularx}{0.48\textwidth}{@{\extracolsep{\fill}}cccccc} \hline\hline
$label$ & $calc.$ & $carrier$ & $\omega_{j}$ & $\Gamma_{j}$ & $S_{j} \times 10^{13}$ \\
& $(\textrm{THz})$ & & $(\textrm{THz})$ & $(\textrm{THz})$ & $(\Omega^{-1}\textrm{cm}^{-1}s^{-1})$ \\ \hline
\raisebox{-.7em}[0pt][0pt]{$\nu_{1}$} & \raisebox{-.7em}[0pt][0pt]{1.20} & $n$ & 1.24 & 0.20 & 6.6 \\
 & & $p$ & 1.19 & 0.20 & 6.4 \\
\raisebox{-.7em}[0pt][0pt]{$\nu_{2}$} & \raisebox{-.7em}[0pt][0pt]{1.62} & $n$ & 1.69 & 0.20 & 4.2 \\
 & & $p$ & 1.67 & 0.17 & 3.4 \\
\hline\hline
\end{tabularx}}
\end{center}
\end{table}

\begin{figure}[!tb]
\includegraphics[width=7.5cm]{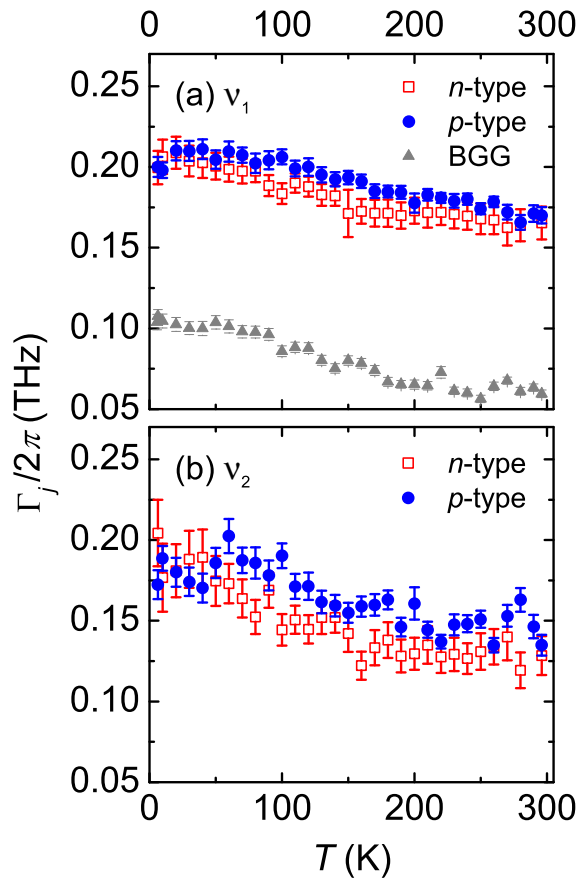}
\caption{(color online)
Temperature dependence of the linewidth (a) $\Gamma_{1}$ for the $\nu_1$ mode and (b) $\Gamma_{2}$ for the $\nu_2$ mode of $\alpha$-BGS. Open squares and filled circles indicate $n$-$\alpha$-BGS and $p$-$\alpha$-BGS, respectively. The linewidth of the lowest-lying mode of Ba$^{2+}$ in BGG is also plotted in (a).
}
\label{fig8}
\end{figure}

\begin{figure}[!tb]
\includegraphics[width=0.48\textwidth ]{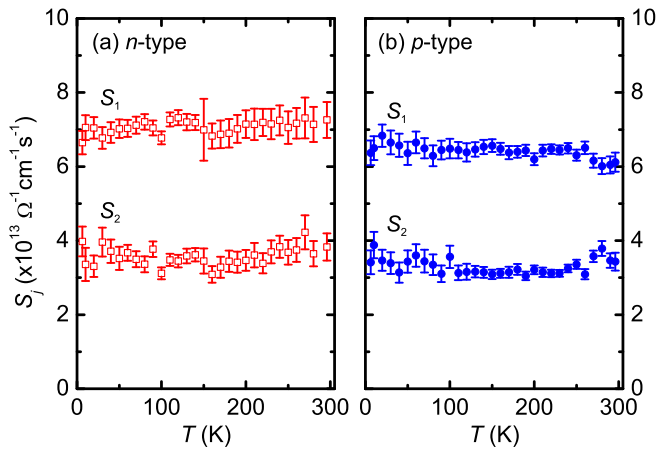}
\caption{(color online)
Temperature dependence of the oscillator strength $S_j$ of the $\nu_j$ mode ($j=1,2$), for (a) $n$-$\alpha$-BGS and (b) $p$-$\alpha$-BGS.
}
\label{fig9}
\end{figure}

Based on the group theory, the irreducible representation of the Ba$^{2+}$ ions' individual vibrating modes in the dodecahedral cages are written as $A_1 + E + T_1 +2T_2$, where two modes with $T_2$ symmetry are infrared active.
The low-lying peaks observed at 1.3 and 1.7\,THz can be assigned as the modes of Ba$^{2+}$ vibrating in the cages within the plane perpendicular to the three-fold axis and along the three-fold axis, labeled $\nu_1$ and $\nu_2$, respectively.
The three fitting parameters for the phonon contributions are plotted as a function of temperature in Figs. \ref{fig7}-\ref{fig9}, and the values at 6\,K are listed in Table \ref{tab1}.
As shown by filled and open squares in Fig. \ref{fig7}, the peak frequencies $\omega_1$ of the $\nu_{1}$ mode coincide at room temperature, and then soften by 7.5\,\% and 11\,\% toward 6\,K for $n$-$\alpha$-BGS and $p$-$\alpha$-BGS, respectively.
The peak frequencies $\omega_2$ of the $\nu_2$ mode, shown as filled and open circles in Fig. \ref{fig7}, are closer and decrease by 2-3\,{\%} upon cooling from room temperature.
However, the temperature dependencies are quite different: for $p$-$\alpha$-BGS, $\omega_2$ decreases monotonically with cooling, whereas for $n$-$\alpha$-BGS, $\omega_2$ shows only slight softening down to 150\,K and then monotonically decreases by 2\,\%.
The temperature dependencies are well described by the model calculations (solid lines) for anharmonic vibrations, \cite{Dahm2007} and are quite similar to those of quasi-on-center Ba$^{2+}$ in type-I BGG. \cite{Mori2009,Iwamoto2013}

Figure \ref{fig8} (a) and (b) show the linewidths $\Gamma _i$ of the $\nu_1$ and $\nu_2$ modes for both $n$- and $p$-$\alpha$-BGS.
The linewidth of the lowest-lying Ba$^{2+}$ mode in BGG\cite{Iwamoto2013} is also shown for comparison.
For both $n$- and $p$-$\alpha$-BGS, $\Gamma_1$ increases by about 18\,{\%}, from 0.17\,THz at room temperature to 0.20\,THz at 6\,K.
As mentioned, such anomalous linewidth broadenings of the lowest-lying modes have also been observed in BGG \cite{Iwamoto2013} and $\beta$-BGS \cite{Mori2011}.
The linewidth $\Gamma_2$ of the $\nu_{2}$ mode also broadens from 0.12-0.13\,THz at room temperature to 0.18-0.20\,THz at 6\,K for $n$- and $p$-$\alpha$-BGS.

Figure \ref{fig9} shows the temperature dependencies of the oscillator strengths $S_1$ and $S_2$.
Both are weakly temperature dependent and coincide well between $n$- and $p$-$\alpha$-BGS, although $S_1$ is slightly smaller for $p$-$\alpha$-BGS.
The observed ratio $S_1/S_2 \simeq 2.0$ is quite consistent with the theoretical value of 2, taking into account the degrees of freedom for the $\nu_1$ and $\nu_2$ modes.
The total contributions $S_1 + S_2$ from Ba$^{2+}$ ions are about $10.0 \times 10^{13}~\Omega ^{-1} {\rm cm}^{-1} s^{-1}$ for $n$-$\alpha$-BGS and about $10.6 \times 10^{13}~\Omega ^{-1} {\rm cm}^{-1} s^{-1}$ for $p$-$\alpha$-BGS, which is 3 times larger than the estimated value of $3.6 \times 10^{13}$ for Ba$^{2+}$ ions, calculated using $M = M_{\rm Ba}$, $N = 8/V_u$ ($V_u = 11.59^3 {\AA}^3$ is the unit cell volume), and $q = 2e$.
These results can be well understood by Eq. \ref{optsumph}, the optical sum rule for the infrared active normal modes.

\section{Discussion}

\subsection{Charge carriers}
\label{sec4-1}
$\chi$ shows no temperature or carrier dependence.
As already mentioned, multiple scattering of charge carriers is taken into account in the DS model \cite{Smith2001} and thus the obtained value of $\chi$, which translates to an effective scattering angle of $116^{\circ}$, suggests the predominance of backscattering by impurities in $\alpha$-BGS.
As also seen in Fig \ref{fig5} (b), $\gamma$ below 50\,K, which is mainly due to impurity scattering, is almost the same for both types of samples.
On the other hand, the carrier density in $n$-$\alpha$-BGS is larger by a factor of 2 than that in $p$-$\alpha$-BGS.
These facts indicate that the concentration of impurities which scatter charge carriers is hardly affected by carrier doping, although charge carriers are doped by a fine tuning of the Ga/Sn relative concentration slightly off stoichiometry.
Here, since the carrier density is calculated as about $10^{-2}$ per unit cell from plasma frequencies, the deviation of the Ga/Sn concentration from the stoichiometry is estimated as about $10^{-2}$ per unit cell, which can be neglected. 
In the framework of $\alpha$-BGS, Ga atoms, whose valence and atomic radius are different from those of Sn, are randomly distributed.
Therefore, we conclude that backscattering in $\alpha$-BGS is mainly caused by the randomly distributed Ga atoms in the framework.
The backscattering of charge carriers by the randomly distributed Ga atoms is proposed as one of the possible candidates for non-Drude like dispersion of optical conductivities in other clathrates, such as $\beta$-BGS \cite{Mori2011} and type-VIII Eu$_8$Ga$_{16}$Ge$_{30}$.\cite{Sichelschmidt2005}

As a temperature rises towards 300\,K, the relaxation rate $\gamma$ of $n$-$\alpha$-BGS increases more rapidly than that of $p$-$\alpha$-BGS.
Since charge carriers interact mainly with the impurities and the thermally excited phonons, such a behavior suggests stronger interactions between phonons and charge carriers in $n$-$\alpha$-BGS than in $p$-$\alpha$BGS.
According to first-principle calculations, \cite{Madsen2003, Kono2010} the valence band consists mainly of Ga and Sn atomic orbitals, whereas the conduction band consists of the atomic orbitals of Ba as well as those of Ga and Sn.
Thus, an important distinction is that electrons induced to the conduction band can interact more strongly with the rattling vibration of Ba than the holes excited to the valence band.

\subsection{Rattling phonons}

The peak frequencies of the $\nu_1$ and $\nu_2$ modes for the Ba$^{2+}$ rattling behave differently between $n$-$\alpha$-BGS and $p$-$\alpha$-BGS, whereas the linewidths of both modes do not show carrier dependence and only slightly increase upon cooling.
The results suggest that doping with different types of charge carriers affects the vibrational states, rather than the dissipative mechanisms.
In $\alpha$-BGS, the Ba$^{2+}$ rattling phonons are expected to interact with the cage acoustic phonons, the charge carriers and impurities.
The disorder in cage potentials caused by the difference in radius between Sn and Ga can also be a relevant factor, as proposed from structural analysis. \cite{Avila2006a}
In the following sections, we will discuss how all these interactions affect the peak frequency and the linewidth.

\subsubsection{Linewidths}

The obtained linewidths in $n$-$\alpha$-BGS and $p$-$\alpha$-BGS agree quantitatively with each other, implying that the density and sign of the charge carriers only weakly affect the linewidth of guest modes at best.
The dissipation by acoustic phonons can be considered negligible since the linewidth does not show any increase upon heating.
As seen in Fig. \ref{fig8} (a), such linewidth broadening toward low temperature is also observed in type-I BGG, which has been attributed to both the strong hybridization between acoustic and rattling phonons and softening due to the anharmonicity. \cite{Iwamoto2013}
Since the linewidths increase by about 0.5\,THz both in BGG and $\alpha$-BGS, we propose the same impurity-scattering model for the hybridized phonon system \cite{Iwamoto2013} to interpret the linewidth broadening in $\hat \sigma$ of $\alpha$-BGS.
The dispersion relation of phonons has not been reported yet for $\alpha$-BGS, but from the sound velocity of $\beta$-BGS which has a similar chemical formula and Debye temperature, \cite{Suekuni2008,Nakayama2011} the frequencies at the zone boundary of the transverse and longitudinal cage acoustic phonons are estimated as about 0.85\,THz and 1.45\,THz in $\beta$-BGS.
Assuming comparable values for $\alpha$-BGS, hybridization of the $\nu_1$ modes with the acoustic branches is expected.
The softening due to anharmonicity increases the phonon DOS at the zone boundary, which produces larger dissipation in the rattling phonon and results in linewidth broadening toward low temperature.

As for the charge carrier contribution, the linewidth of the $\nu_1$ mode in $\alpha$-BGS is larger by about 0.1\,THz than that in BGG, irrespective of temperature or the type of charge carriers.
Such an additional linewidth must result from electron-phonon interactions and/or structural disorder.
The linewidth $\Gamma _{i,ep}$ from electron-phonon coupling for the $\nu_i$ mode is expressed as $\Gamma _{i,ep} = (\pi/3) \lambda _{i,ep} \omega_i^2 N_F$, where $N_F$ is the electronic density-of-states at the Fermi energy and $\lambda _{i,ep}$ is the phonon Gr{\"u}neisen parameter for the $\nu_i$ mode. \cite{Hattori2007,Grimvall1981}
Here, $\lambda _{i,ep}$ is related to the relaxation rate $\gamma_{ep}$ of charge carriers by the Bloch-Gr{\"u}neisen formula for Einstein phonons. \cite{Grimvall1981,Lortz2008}
$N_F$ is estimated as 1.39 states/eV in $n$-$\alpha$-BGS and 0.96 states/eV in $p$-$\alpha$-BGS, respectively, by using the plasma frequency $\omega _p$ at 6\,K and assuming a free-carrier model.
If the additional linewidth is caused by the interaction with charge carriers, $\lambda _{1,ep}$ is roughly estimated as 10.
However, when the obtained value of $\lambda _{1,ep}$ is substituted into the Bloch-Gr{\"u}neisen formula, the contribution from the $\nu_1$ mode is estimated as 40-50\,THz, larger by a factor of about 10-20 than the experimental value $\gamma$.
In other words, the linewidth induced by the interaction of Ba$^{2+}$ rattling with charge carriers should be of order 0.01\,THz at most.

Thus, the remaining possibility should reside in the structural disorder due to Ga atoms.
As revealed by EXAFS analysis, \cite{Jiang2008} the bonding lengths of Ge and Ga are almost the same in BGG, but in $\alpha$-BGS the atomic radius of Sn is significantly larger than that of Ga. \cite{Avila2006b}
Since the sites on the framework are randomly occupied by Ga atoms, the difference in atomic radius can increase the linewidth through the deformation of the cage potential.
The difference in the number of Ga atoms between $n$- and $p$-$\alpha$-BGS is negligibly small as already discussed in \S \ref{sec4-1}.
Therefore, the structural disorder is the likely candidate to induce additional linewidth in the phonon spectra.

We also note that the phonon spectra of $\alpha$-BGS are well reproduced by the Lorentz model.
This is in clear contrast to the phonon spectra of BGG, which consist of a sharp Lorentzian peak superimposed on a broad tail weighted in the lower frequency regime as previously reported. \cite{Iwamoto2013}
According to the impurity-scattering model, the sharp peak and tail structure in BGG is caused by a rapid change of the phonon density-of-states (DOS) around the hybridization gap. \cite{Iwamoto2013}
As discussed above, in $\alpha$-BGS, the structural disorder and/or the electron-phonon interaction contribute to the linewidth of rattling phonons.
These additional contributions fill the would-be hybridization gap, smoothing the phonon DOS, and the rattling phonon spectra of $\alpha$-BGS have no such clear sideband structures reflecting the phonon DOS.

\subsubsection{Peak Frequency}

\begin{table}
\caption{
Calculated phonon frequencies and fitting parameters of $\alpha$-BGS at 6\,K.
}
\label{tab2}
\begin{center}
{\renewcommand\arraystretch{1.2}
\setlength{\arrayrulewidth}{0.2pt}
\begin{tabularx}{0.48\textwidth}{@{\extracolsep{\fill}}ccccc} \hline\hline
$label$ & $carrier$ & $\omega_{0}~(\textrm{THz})$ & $\bar b$ \\ \hline
\raisebox{-.7em}[0pt][0pt]{$\nu_{1}$} & $n$ & 1.23 & $1.60 \times 10^{-2}$ \\
 & $p$ & 1.17 & $2.59 \times 10^{-2}$ \\
\raisebox{-.7em}[0pt][0pt]{$\nu_{2}$} & $n$ & 1.68 & $3.9 \times 10^{-3}$ \\
 & $p$ & 1.66 & $6.8 \times 10^{-3}$ \\
\hline\hline
\end{tabularx}}
\end{center}
\end{table}

The softening of the rattling vibration observed in Fig. \ref{fig7} is well understood by assuming an anharmonic potential which includes a quartic term in the guest ion displacement.
When the center of mass coordinate $\vec X$ of the cage and the coordinate of the guest ion $\vec x$ are considered, the potential in the cage is applied to the relative coordinate $\vec r = \vec x - \vec X$
\begin{equation}
V(\vec r) = \sum_{i =x,y,z} \left( \frac{1}{2} k_i r_i^2 + \frac{1}{4} b _i r_i^4 \right) .
\label{poten}
\end{equation}
As already mentioned, since the rattling excitation may have a more propagating nature through the interaction with acoustic phonons, \cite{Iwamoto2013} the softening of the rattling excitation is well reproduced by the mean field approximation \cite{Dahm2007},
\begin{equation}
\omega _{MF}^2 \simeq k_x/m + 3 ( b_x /m ) \langle r_x^2 \rangle ,
\label{meanf}
\end{equation}
where $m$ is the reduced mass of the cage and the guest and $\langle r_x^2 \rangle$ is the mean square value of $r_x$.
To obtain the anharmonicity parameters of the rattling phonons, we carried out a fitting of Eq. \ref{meanf} to $\omega _1$ and $\omega _2$.
The solid lines in Fig. \ref{fig7} are the fitted curves, and the parameters $\omega _0$ and $\bar b$ are listed in Table \ref{tab2}, where {$\omega _0 = \sqrt{k_x/m}$} is the peak frequency and {$\bar b = \hbar b_x /(m^2 \omega _0^3)$} is a dimensionless anharmonicity parameter.
The Table values show that $\bar b$ in $n$-$\alpha$-BGS are smaller by 40\,{\%} than those in $p$-$\alpha$-BGS, implying in weaker anharmonicity of the Ba$^{2+}$ rattling in $n$-$\alpha$-BGS.
According to our discussions in the previous section, the elastic scattering by the local impurities and structural disorder are not affected by the carrier doping, so the anharmonicity parameters should not be modified significantly by these effects.
To understand the carrier dependence of $\bar b$ in $\alpha$-BGS, the interactions with conducting and/or localized electrons should be taken into account.
As discussed in \S \ref{sec4-1}, the temperature variation of $\gamma$ indicates that the electron-phonon interaction in $n$-$\alpha$-BGS
is much stronger than that in $p$-$\alpha$-BGS.
Thus, carrier dependence in the peak frequencies may also be caused by the electronic structure
where the Ba atomic orbitals hybridize with those of Ga and Sn in the conduction band. \cite{Kono2010}
Electrons on Ba$^{2+}$ ions can vary the effective charge of Ba$^{2+}$ and hence modify the potential in the atomic cage.
The frequency and/or anharmonic potential may be renormalized by such an interaction with $n$-type carriers, resulting in the carrier dependence of $\bar b$ and $\omega _0$.

As discussed above, the charge carriers on Ba$^{2+}$ ions and the rattling phonons are considered to interact with each other, which is a cause of the carrier dependence observed in optical conductivity spectra.
Then, it is expected that the interaction between charge carriers and phonons affect the thermal transport. 
In terms of the unusual lattice thermal conductivity in these materials, there are two noteworthy candidate mechanisms that can lead to charge carrier dependence of $\kappa_L$: (1) the interaction between rattling and acoustic phonons and (2) the direct interaction between phonons and charge carriers.
The present results imply that a renormalization of the guest potential occurs in the atomic cage, which result from the existence of Ba orbitals in the conduction band, and the consequent presence of doped electrons on the Ba$^{2+}$ ions. 
Carrier dependence of the former interaction may arise from such a renormalization of the guest potential. 
In other words, an enhancement of the interaction between rattling and acoustic phonons in $p$-type materials may be proposed due to stronger anharmonicity.
As for the latter, it may also be proposed that the charge carriers on the framework interact directly with acoustic phonons and with the rattling phonon through charging of cages, and then contribute to reduce the lattice thermal conductivity in $p$-$\alpha$-BGS, since $p$-type carriers are induced onto the valence band which consists mainly of Ga and Sn atomic orbitals. \cite{Madsen2003, Kono2010}
The relative influence of each mechanism to the carrier dependence in the lattice thermal conductivity is still an open question.
However, our analysis establishes that the charge carriers do play a significant role in both mechanisms proposed to understand the dynamical properties of phonons and hence the thermal properties of clathrate compounds.

\section{Conclusion}

We have clarified the dynamical motion of the rattling phonons and charge carriers in $\alpha$-BGS single crystals with both $n$- and $p$-type carriers by THz-TDS.
The optical conductivity spectra are well reproduced by superpositions of contributions from valence electrons, conducting carriers, and infrared-active optical phonons.
The contribution from charge carriers is well reproduced by the Drude-Smith model, in which multiple scattering of charge carriers is taken into account.
The relaxation rate behaves differently depending on the charge carrier sign, suggesting a stronger electron-phonon interaction in $n$-$\alpha$-BGS than in $p$-$\alpha$-BGS.
Ba$^{2+}$ rattling modes in the oversized cages show softening and broadening towards low temperature.
While the linewidths agree quantitatively between $n$- and $p$-$\alpha$-BGS, the peak frequency shows clear carrier dependence.
The linewidths are found to increase due to impurities and structural disorder, rather than interaction with conducting carriers, due to the low carrier density of $10^{18}$-$10^{19}$\,cm$^{-3}$.
On the other hand, the anharmonic potential of Ba$^{2+}$ in $n$-$\alpha$-BGS seems modified by electrons in the conduction band,
which consists of Ba atomic orbitals as well as Ga and Sn orbitals.

The present experiments have indicated that the interaction between rattling phonons and charge carriers contributes not only to the scattering of charge carriers, but also to the vibrational states of rattling phonons, particularly in the $n$-type material.
As also suggested in Ref. \cite{Bentien2006}, the interaction with charge carriers may also affect the dynamical properties of the acoustic phonons directly.
The obtained results also give rise to an interesting problem for the compounds with {\it off-centered} guest ions, {\it i. e.}, $\beta$-BGS and type-I Eu$_8$Ga$_{16}$Ge$_{30}$. \cite{Sales2001}
In these materials the guest ion vibrates in a multi-well potential and is expected to move among each potential minimum by quantum tunneling.
It is theoretically proposed that the interaction between the conducting electrons and the tunneling states of off-center rattling systems results in a multi-channel Kondo effect. \cite{Hattori2005,Hotta2007}
The detailed interactions among the charge densities at the guest ion sites, the cage phonons, and the {\it off-center} rattling remains as an interesting open problem in these cagelike materials.
From the present studies, such interactions are expected to contribute not only to the lifetime of phonons but also to a renormalization of phonon frequencies.
The dynamical properties of low-energy phonons need to be investigated in clathrate compounds through Brillouin scattering and inelastic neutron and X-ray scattering.

\section*{ACKNOWLEDGMENTS}

K.I. is supported financially by the Global COE program ``Materials Integrations'', Tohoku University, and one of us (H.M.) partially by the Grant-in-Aid for Scientific Research (C) 23500056. The works at Sendai have been supported by Grants-in-Aid for Scientific Research (A)(15201019) and Young Scientists (B)(24740194), the priority area ``Nanospace'' from MEXT, Japan, and Murata Science Foundation. The works at Higashi Hiroshima have been supported by Grants-in-Aid for Scientific Research (A)(18204032), the priority area ``Nanospace''(19051001) and ``Skutterudite''(15072205), and the innovative areas ``Heavy Electrons'' (20102004) from MEXT, Japan, and the Sasakawa Scientific Research from Japan Science Society. MAA also thanks the financial support of FAPESP, No. 2012/17562-9.


\begin{thebibliography}{99}

\bibitem{Nolas1998} G. S. Nolas, J. L. Cohn, G. A. Slack and S. B. Schujman, Appl. Phys.Lett. {\bf 73}, 179, (1998).
\bibitem{Cohn1999} J. L. Cohn, G. S. Nolas, V. Fessatidis, T. H. Metcalf, G. A. Slack, Phys. Rev. Lett. {\bf 82}, 779, (1999).
\bibitem{Sales2001} B. C. Sales, B. C. Chakoumakos, R. Jin, J. R. Thompson, and D. Mandrus, Phys. Rev. B, {\bf 63}, 245113, (2001).
\bibitem{Christensen2008} M. Christensen, A. B. Abrahamsen, N. B. Christensen, F. Juranyi, N. H. Andersen, K. Lefmann, J. Andreasson, C. R. H. Bahl, and B. B. Iversen, Nat. Mater., {\bf 7}, 811, (2008).
\bibitem{Eisen1986} B. Eisenmann, H. Schafer, and R. Zahler, J. Less-Common Met., {\bf 118}, 43 (1986).
\bibitem{Huo2005} D. Huo, T. Sakata, T. Sasakawa, M. A. Avila, M. Tsubota, F. Iga, H. Fukuoka, S. Yamanaka, S. Aoyagi, and T. Takabatake, Phys. Rev. B, {\bf 71}, 075113, (2005).
\bibitem{Avila2006c} M. A. Avila, D. Huo, T. Sakata, K. Suekuni, and T. Takabatake, J. Phys.: Condens. Matter, {\bf 18}, 1585, (2006).
\bibitem{Avila2006b} M. A. Avila, K. Suekuni, K. Umeo, H. Fukuoka, S. Yamanaka, and T. Takabatake, Phys. Rev. B, {\bf 74}, 125109, (2006).
\bibitem{Suekuni2008} K. Suekuni, M. A. Avila, K. Umeo, H. Fukuoka, S. Yamanaka, T. Nakagawa, and T. Takabatake, Phys. Rev. B {\bf 77}, 235119 (2008).
\bibitem{Avila2006a} M. A. Avila, K. Suekuni, K. Umeo, and T. Takabatake, Physica B, {\bf 383}, 124, (2006).
\bibitem{Avila2008} M. A. Avila, K. Suekuni, K. Umeo, H. Fukuoka, S. Yamanaka, and T. Takabatake, Appl. Phys. Lett. {\bf 92}, 041901 (2008).
\bibitem{Bentien2004} {A. Bentien, M. Christensen, J. D. Bryan, A. Sanchez, S. Paschen, F. Steglich, G. D. Stucky, and B. B. Iversen, Phys. Rev. B {\bf 69}, 045107, (2004).}
\bibitem{Bentien2006} A. Bentien, S. Johnsen, and B. B. Iversen, Phys. Rev. B, {\bf 73}, 094301, (2006). 
\bibitem{Mori2009} T. Mori, S. Goshima, K. Iwamoto, S. Kushibiki, H. Matsumoto, N. Toyota, K. Suekuni, M. A. Avila, T. Takabatake, T. Hasegawa, N. Ogita, and M. Udagawa, Phys. Rev. B {\bf 79}, 212301 (2009).
\bibitem{Iwamoto2013} K. Iwamoto, S. Kushibiki, H. Honda, S. Kjitani, T. Mori, H. Matsumoto, N. Toyota, K. Suekuni, M. A. Avila, and T. Takabatake, J. Phys. Soc. Jpn., {\bf 82}, 024601 (2013).
\bibitem{Dahm2007} T. Dahm and K. Ueda, Phys. Rev. Lett. {\bf 99}, 187003, (2007).
\bibitem{Mori2011} T. Mori, K. Iwamoto, S. Kushibiki, H. Honda, H. Matsumoto, N. Toyota, M. A. Avila, K. Suekuni, and T. Takabatake, Phys. Rev. Lett., {\bf 106}, 015501, (2011).
\bibitem{Matsumoto2009} H. Matsumoto, T. Mori, K. Iwamoto, S. Goshima, S. Kushibiki, and N. Toyota, Phys. Rev. B {\bf 79}, 214306 (2009).
\bibitem{Nakayama2011} T. Nakayama and E. Kaneshita, J. Phys. Soc. Jpn. {\bf 80}, 104604, (2011).
\bibitem{Sichelschmidt2005} J. Sichelschmidt, V. Voevodin, V. Pacheco, Yu. Grin, F. Steglich, T. Nishi, and S. Kimura, Eur. Phys. J. B, {\bf 46}, 363, (2005).
\bibitem{Tou2011} H. Tou, K. Sonoda, Y. Nishikawa, H. Kotegawa, H. Suekuni, T. Onimaru, and T. Takabatake, J. Phys. Soc. Jpn., {\bf 80} SA039 (2011).
\bibitem{Ishii2005} I. Ishii, H. Higaki, T. Sakata, D. Huo, T. Takabatake, and T. Suzuki, Physica B, {\bf 359-361} 1210 (2005).
\bibitem{Mori2008} T. Mori, E. J. Nicol, S. Shiizuka, K. Kuniyasu, T. Nojima, N. Toyota, and J. P. Carbotte, Phys. Rev. B {\bf 77}, 174515 (2008).
\bibitem{Dressel} {M. Dressel and G. Gr\"{u}ner, {\it Electrodynamics of Solids} (Cambridge, 2002).}
\bibitem{Smith2001} N. V. Smith, Phys. Rev. B, {\bf 64}, 155106, (2001).
\bibitem{Madsen2003} {G. K. H. Madsen, K. Schwarz, P. Blaha, and D. J. Singh, Phys. Rev. B, {\bf 68}, 125212, (2003).} 
\bibitem{Kono2010} Y. Kono, N. Ohya, T. Taguchi, K. Suekuni, T. Takabatake, S. Yamamoto, and K. Akai, J. Appl. Phys. {\bf 107}, 123720 (2010).
\bibitem{Hattori2007} K. Hattori and K. Miyake, J. Phys. Soc. Jpn. {\bf 76}, 094603, (2007).
\bibitem{Grimvall1981} G. Grimvall, {\it The Electron-Phonon Interaction in Metals} (North-Holland, Amsterdam, 1981).
\bibitem{Lortz2008} R. Lortz, R. Viennois, A. Petrovic, Y. Wang, P. Toulemonde, C. Meingast, M. M. Koza, H. Mutka, A. Bossak, and A. S. Miguel, Phys. Rev. B {\bf 77}, 224507, (2008).
\bibitem{Jiang2008} Y. Jiang, F. Bridges, M. A. Avila, T. Takabatake, J. Guzman, and G. Kurczveil, Phys. Rev. B {\bf 78} 014111 (2008).
\bibitem{Hattori2005} K. Hattori, Y. Hirayama, and K. Miyake, J. Phys. Soc. Jpn. {\bf 74} 3306 (2005).
\bibitem{Hotta2007} T. Hotta, J. Phys. Soc. Jpn. {\bf 76} 084702 (2007).

\end{thebibliography}
\end{document}